\documentclass{article}

\usepackage{neurips_2022}

\usepackage{natbib}
\setcitestyle{numbers,square}
\usepackage[utf8]{inputenc} 
\usepackage[T1]{fontenc}    
\usepackage{hyperref}       
\usepackage{url}            
\usepackage{booktabs}       
\usepackage{amsfonts}       
\usepackage{nicefrac}       
\usepackage{microtype}      
\usepackage{xcolor}         
\usepackage{graphicx}
\hypersetup{hidelinks}
\title{Brain Cortical Functional Gradients Predict Cortical Folding Patterns via Attention Mesh Convolution}

\author{%
  Li Yang \\
  School of Automation\\
  Northwestern Polytechnical University\\
  Xi’an, China 710072 \\
  \texttt{lyang2021@mail.nwpu.edu.cn} \\
   \AND
   Zhibin He \\
   School of Automation\\    
   Northwestern Polytechnical University \\
   \texttt{hezhibin@mail.nwpu.edu.cn} \\
   \And
   Changhe Li \\
   School of Automation\\     
   Northwestern Polytechnical University \\
   \texttt{changheli@mail.nwpu.edu.cn} \\
   \And
   Junwei Han \\
   School of Automation\\ 
   Northwestern Polytechnical University \\
   \texttt{junweihan2010@gmail.com} \\
   \And
   Dajiang Zhu \\
   University of Texas at Arlington \\
   \texttt{dajiang.zhu@uta.edu} \\
   \And
   Tianming Liu \\
   University of Georgia \\
   \texttt{tianming.liu@gmail.com} \\
   \And
   Tuo Zhang \\
   School of Automation \\
   Northwestern Polytechnical University \\
   \texttt{tuozhang@nwpu.edu.cn} \\
}

\begin{document}

\maketitle

\begin{abstract}
  A close relation between brain function and cortical folding has been demonstrated by macro-/micro- imaging, computational modeling, and genetics. Since gyri and sulci, two basic anatomical building blocks of cortical folding patterns, were suggested to bear different functional roles, a precise mapping from brain function to gyro-sulcal patterns can provide profound insights into both biological and artificial neural networks. However, there lacks a generic theory and effective computational model so far, due to the highly nonlinear relation between them, huge inter-individual variabilities and a sophisticated description of brain function regions/networks distribution as mosaics, such that spatial patterning of them has not been considered. To this end, as a preliminary effort, we adopted brain functional gradients derived from resting-state fMRI to embed the "gradual" change of functional connectivity patterns, and developed a novel attention mesh convolution model to predict cortical gyro-sulcal segmentation maps on individual brains. The convolution on mesh considers the spatial organization of functional gradients and folding patterns on a cortical sheet and the newly designed channel attention block enhances the interpretability of the 
 contribution of different functional gradients to cortical folding prediction. Experiments show that the prediction performance via our model outperforms other state-of-the-art models. In addition, we found that the dominant functional gradients contribute less to folding prediction. On the activation maps of the last layer, some well-studied cortical landmarks are found on the borders of, rather than within, the highly activated regions. These results and findings suggest that a specifically designed artificial neural network can improve the precision of the mapping between brain functions and cortical folding patterns, and can provide valuable insight of brain anatomy-function relation for neuroscience. 
\end{abstract}

\section{Introduction}

Relation between brain function and cortical folding has long been intriguing due to its importance in offering profound insight into brain functional operation mechanisms, brain development and evolution, as well as being of clinical applications \cite{troiani2020use}. Gyri and sulci, the convex and concave cortical folding patterns, are two basic anatomical units of cortical morphology. It has been found that gyri/sulci align well with cortical functional boundaries \cite{fischl2008cortical}, at least in primary functional cortex. Specific gyro-sulcal patterns, such as the $plis$ $de$ $passage$ gyri \cite{regis2005sulcal}, topology of local gyri and sulci (such as H-shape sulci in frontal lobe \cite{troiani2020use}, H-shape gyri in fusiform gyri \cite{cachia2018interindividual} and the ventromedial prefrontal cortex (vmPFC) in DMN region \cite{lopez2019human}), were suggested to correspond to subtypes of a specific function (such as cognitive and perception ability associated with frontal lobe and reading ability associated with fusiform gyri) or relate to the placement of a specific functional region (such as functional activation location in vmPFC). In \cite{liu2019cerebral}, functional MRI signals were suggested to have stronger low frequency components on gyri, whereas sulci have signals of stronger high frequency components. Neuronal difference in their distributions \cite{hilgetag2005developmental}, types \cite{rash2019gliogenesis}, wiring patterns \cite{xu2010axons} and even morphology \cite{hilgetag2005developmental} were reported to be present between gyri and sulci, which was further suggested to lead to differentiated cell-level functioning patterns. In addition, expression of a huge number of genes was reported to be different between gyri and sulci \cite{de2015discrete}. Among them, interestingly, the same group of genes can lead to the formation of gyri/sulci on gyrencephalic species while mediating brain function distribution patterns on lissencephaly species \cite{de2015discrete}, suggesting a possible deeply rooted bond between cortical folding pattern and brain functions. Importantly, the functional differentiation between gyri and sulci could also provide clues to artificial neural network design. For example, gyri were suggested to serve as functional hubs and sulci as local processors, whereas sulci were isolated by gyri on the cortical sheet 
\cite{deng2014functional}. Recalling that graph architecture of an artificial neural network in the sweet spot resembles the real neural networks \cite{you2020graph}, such a gyro-sulcal (hub-nonhub) spatial distribution might suggest an alternative approach for neural network design.

In spite of the importance in anatomy-function relation, no explicit explanation or generic theory of such a relation can be satisfactorily applied across brain regions and species \cite{fischl2008cortical}\cite{frost2012measuring}. Also, very few computational models are found \cite{liu2019cerebral} to provide a precise prediction from brain function to cortical folding. The reasons could be attributed to many folds: 1) the mapping between brain function and folding pattern is far from linear; 2) inter-individual variabilities are huge in the layout of both gyro-sulcal patterns and functional regions \cite{fischl2008cortical}\cite{frost2012measuring}; 3) brain function was usually delineated by atlases where “abrupt” boundaries were used to segregate functional “units” or “networks”, the functional attributes within which were considered to be of the same. However, the transition of a brain function from one to another was suggested to be “smooth” and “gradual”, which also gains supports from the related genetic expression \cite{huntenburg2018large}; 4) the spatial patterning of brain function distribution was not fully used to predict cortical folding.

To answer the aforementioned questions, as a preliminary effort, we proposed to used artificial neuronal networks to investigate the predictive ability of brain functional gradients to cortical folding patterns at macro scale. By means of resting-state fMRI (rsfMRI) and structural MRI (sMRI) in Human Connectome Project (HCP) dataset \cite{van2013wu}, we estimated the global functional gradients, embedding axes that encode the “gradual” differences in mesh vertices' connectivity patterns and can be present on a cortical sheet (top-left in Figure 1), where the prediction of gyro-sulcal segmentation (top-right in Figure 1) was performed. Technically, we proposed to use a mesh convolution model\cite{hu2022subdivision} based on the classic U-net architecture, which has been widely and successfully applied to medical image segmentation and is a close analogy with our 
task. In addition, to enhance the interpretability of the mapping from functional gradients to cortical folding maps, we added a channel attention block \cite{hu2018squeeze} to the head of the U-net mesh convolution model, where functional gradients, as multi-channels, were re-weighted by the attention block before being fed to mesh convolution. The weights can encode the contribution of functional gradients to cortical folding prediction. Overall, the main contributions of this work are:
\begin{itemize}
    \item We proposed a new concept of adopting functional gradients to encode functional connectivities and present them on the cortex, such that mesh convolution can be adopted to predict gyri/sulci from spatial patterns of functional connectivity embedding.
    \item Channel attention was designed to enhance the interpretability of the nonlinear mapping between brain function and cortical folding.
    \item We explored and defined the activation maps of different layers and found novel clues to function-anatomy relation, which is valuable for both real and artificial neural networks.
\end{itemize}

\section{Related Works}
\label{gen_inst}

\subsection{Mesh Convolution}
Meshes are composed of three distinct types of geometric primitives: vertices, edges, and faces. The classification on meshes is a basic research area which has made great progress with the development of deep learning. DiffusionNet \cite{sharp2022diffusionnet} introduces a general-purpose approach to deep learning on 3D surfaces. The networks can be discretized on various geometric representations such as triangle meshes or point clouds, and can even be trained on one representation and then applied to another. HodgeNet \cite{smirnov2021hodgenet} uses many features that make it an attractive alternative for learning from meshes. During inference, its structure resembles that of most spectral geometry processing algorithms: construct a useful operator, and compute features from its spectrum. However, the above two approaches, as well as many other methods, including Pointnet \cite{qi2017pointnet}, MeshCNN  \cite{hanocka2019meshcnn}, PD-MeshNet \cite{milano2020primal}, MeshWalker \cite{lahav2020meshwalker}, PFCNN \cite{yang2020pfcnn}, need geometric features of meshes, such as coordinates of vertices, which is not applicable to our task that learns and predicts "texture" features on meshes. BrainSurfCNN \cite{ngo2020connectomic} and SubdivNet \cite{hu2022subdivision} provides a solution to the problem. BrainSurfCNN is a method which uses spherical convolutional kernel \cite{jiang2019spherical} to predict the task fMRI contrasts from resting-state functional connectivity ones under a regression framework. The work in \cite{jiang2019spherical} uses only one-ring neighborhood for convolution, which limits the extraction of long-range features. SubdivNet \cite{hu2022subdivision} uses a regular and uniform downsampling scheme to establish a fine-to-coarse mesh hierarchy. The convolutions efficiently support stride and large dilation, allowing the model to better capture long-range features. At the same time, the SubdivNet is a general approach, in which one can easily use 3D CNN on the mesh as similar as using the 2D CNN on the image without geometric features. In our method, we choose the convolution operation of SubdivNet to learn the "texture" features, rather than geometric features, for the purpose of predicting cortical folding patterns from brain cortical functional gradients.

\subsection{Attention Mechanisms}
Attention mechanisms were introduced into computer vision with the aim of effectively finding the salient regions by imitating the human visual system. Recently, a large number of attention mechanisms and approaches for deep learning are proposed to improve the performance of classification. Channel attention adaptively recalibrates the weight of each channel, and determines what to pay attention to. Hu and colleagues \cite{hu2018squeeze} proposed the concept of channel attention and presented SENet for this purpose. The core of SENet is a squeeze-and-excitation (SE) block which is used to collect global information, capture channel-wise relationships and improve representation ability. An SE block can be added after each residual unit due to their low computational resource requirements, which is suitable for our model. Wang and colleagues \cite{2019ECA} proposed the efficient channel attention (ECA) block which uses a 1D convolution to determine the interaction between channels, instead of dimensionality reduction, which can directly model correspondence between weight vectors and inputs. However, it is noticed that the ECA block is not very suitable for models with a small number of channels 
in our experiments (as demonstrated in the Table 1 and Figure 2). \cite{ferrari2018computer} proposed the convolutional block attention module (CBAM) which stacks channel attention and spatial attention in series. However, the method of spatial attention which only fits 2D image can not be directly used in mesh. As a result, the SEnet was chose in our work.

\subsection{Predicting Gyro-sulcus From fMRI}
So far, there are very few works in predicting cortical folding pattern from fMRI via a deep learning model. Liu and colleagues \cite{liu2019cerebral} designed a convolutional neural network (CNN) based classifier, which can differentiate gyral and sulcal fMRI signals with a reasonable accuracy (0.67 on resting state fMRI). However, in this model, vertices on surfaces are samples (with a gyrus/sulcus label) independent from each other, whereas functional regions are not spatially isolated and independent. Either functional network studies or functional gradient ones demonstrated that remote or neighboring regions cooperate with each other to elicit and maintain a brain function \cite{huntenburg2018large}\cite{damoiseaux2006consistent}. Therefore, functional connectivity induced gradients that encode the spatial embedding of brain neuronal interactivity could be more suitable for prediction of spatial layout of gyri/sulci. Also, the relatively lower classification accuracy in \cite{liu2019cerebral} could be ascribed to the direct use of noisy fMRI signals as predictive features, whereas connectivities could mitigate such side effects and direct the attention to brain functions at a global scale rather than a vertex-wise one. Zhao and colleagues \cite{zhao2021exploring} proposed a novel Hierarchical Interpretable Autoencoder (HIAE) to explore the hierarchical functional difference between gyri and sulci. The hierarchical features learned by autoencoder can be embedded into a one-dimensional vector which interprets the features as spatial-temporal patterns. However, its major aim is embedding hierarchical and spatial-temporal functional features into a one-dimensional vector, whereas the functional difference between gyri and sulci was only a \emph{post hoc} analysis.

\begin{figure}[htp]
  \centering
  \includegraphics[width=14cm]{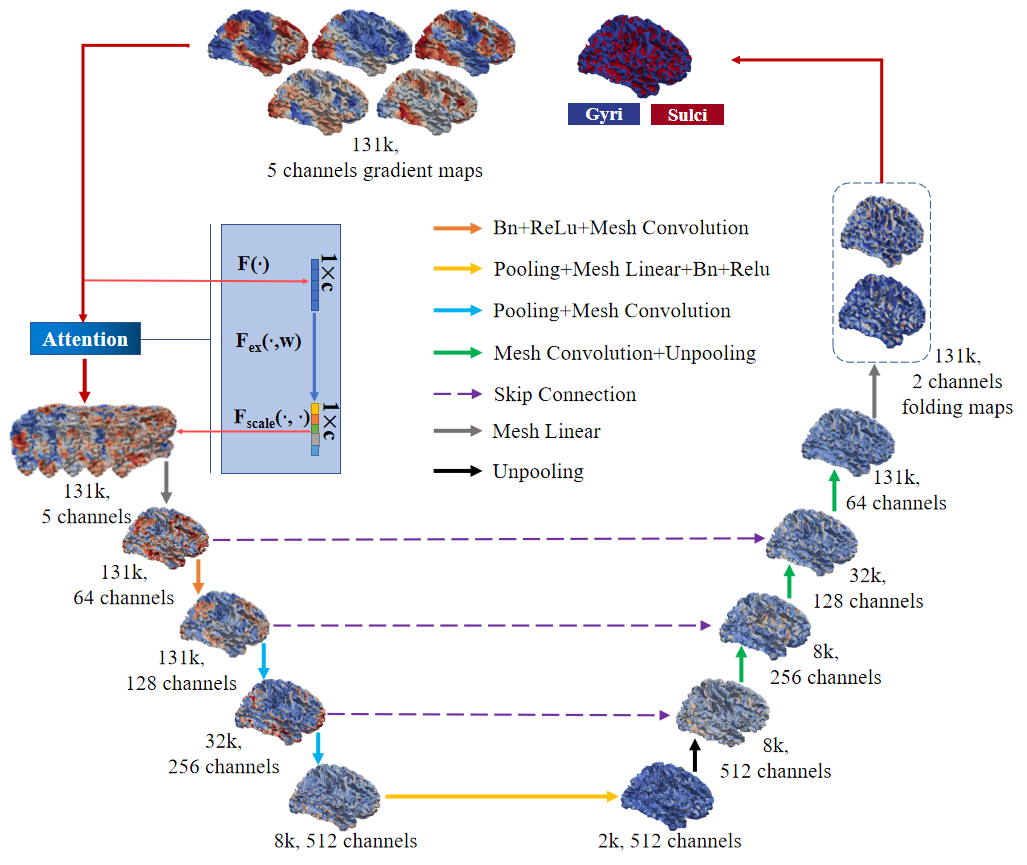}
  \caption{Attention mesh convolution network architecture}
\end{figure}
\section{Method}
\label{headings}

In general, we designed a novel attention mesh convolution method based on U-net architecture to predict the gyro-sulcal segmentation map from brain functional gradient maps. The model’s inputs are multi-channel functional gradients on icosahedral mesh \cite{baumgardner1985icosahedral}, while its output is a two-channel binary map on the same icosahedral mesh (Figure 1).

\subsection{Functional Gradients and Cortical Folding Patterns}

In brief, for each subject, a functional connectivity matrix was calculated using the correlation coefficient for all pairs between vertices (32,492 per hemisphere) on HCP grayordinate mesh system, introduced in section 4.1. For each row of this “dense” functional connectome matrix, the values of the top 10\% of connections were retained, whereas all others were zeroed. The negative values in the remaining connections were zeroed as well. The similarity of the connectivity matrix $\mathbf{A}$ was ensured by calculating the cosine distance between all pairs of rows, yielding a symmetric affinity matrix $\mathbf{L}$. On this matrix, diffusion embedding \cite{margulies2016situating} that allows local and long distance connections to be more effectively projected into a common space, was employed to yield unitless components, also known as functional “gradients”. The positions of nodes on these embedding axes encode the differences in nodes’ connectivity patterns. In this work, the number of gradients was set to 5. Results of other numbers of gradients were compared. It is noted that each functional gradient is associated with an eigen value that suggests its dominance. It has been demonstrated that gradients decomposed from different subjects are consistent, whereas their dominance could vary across subjects \cite{vos2020brainspace}. Hence, the algorithm in \cite{vos2020brainspace} used the group-mean matrix to yield the group-wise gradients, that were further used as templates to identify the cross-subject correspondence of gradients from individuals. The top-left panel of Figure 1 illustrates the aligned 5 functional gradient maps from one subject. Surface mean curvature \emph{k} was used to segment surface to gyri (\emph{k}>0) and sulci (\emph{k}<0) (the right-top panel of Figure 1).

\subsection{Channel Attention Block}

In order to boost the representation power of a network, the channel attention block SEnet \cite{hu2018squeeze} was applied in front of mesh convolution. In this attention block, a self-gating mechanism based on channel dependence yields the excitation of each channel (functional gradient in this work), such that channels of input were then re-weighted and fed to the following layers. This weight vector $\mathrm{w}$ could encodes the respective contribution of functional gradients to folding pattern prediction.

As shown in Figure 1 and Equation(1\&2), the $\mathbf{F}$ $(\cdot)$ is global average pooling function which compresses the characteristics of each channel into a real number, where $\mathrm{c}$ denotes the channel number, $N_{f}$ denotes the number of mesh faces, $G_{c}$ denotes the channel's gradient contrasts. Then, a $\mathrm{1*c}$ vector was sent to $\mathbf{F}_{\mathrm{ex}}(\cdot, \mathrm{w})$ which was implemented by a fully connected layer. In our method, we encoded 5 channels to 2 channels and then decoded them back to 5, in which $\mathrm{w}$ is a learnable parameter. After these operations, $\mathrm{w}$ gained the weight for every input channel, and then $\mathbf{F}_{\mathrm{scale}}(^\cdot, ^\cdot)$ scaled the $\mathrm{1*c}$ vector and make the gradient contrasts on every mesh face multiplied by the corresponding channel weight. Finally, the re-weighted channels were fed to the mesh U-Net.
\begin{equation}\label{eq1}  
z = \mathbf{F}\left(\mathbf{G}_{c}\right)=\frac{1}{N_f} \sum_{i=1}^{N_f}  G_{c}(i)
\end{equation}
\begin{equation}\label{eq2}  
\mathbf{s}=\mathbf{F}_{e x}(\mathbf{z}, \mathbf{W})=\sigma(g(\mathbf{z}, \mathbf{W}))=\sigma\left(\mathbf{W}_{2} \delta\left(\mathbf{W}_{1} \mathbf{z}\right)\right)
\end{equation}

\subsection{Mesh CNN with U-net Architecture and Loss Function}

The model is based on the U-Net architecture \cite{ronneberger2015u} using the mesh convolutional kernel proposed in \cite{hu2022subdivision}. As shown in Figure 1, after the attention block, a 'meshlinear' operation was used to generate 64 channels. This operation is a linear operation on meshes which resembles '1D' convolution on images. Then, the results above were sent to the classic U-Net framework. The kernel size was set to 3 on all mesh convolution operations, which was designed in \cite{hu2022subdivision}, and the 'mean pooling' was used for downsampling. A difference between the classic U-Net framework and ours is that a 'meshlinear' operation was used after the third 'pooling' operation (yellow arrow in Figure 1), and we found this change improves the accuracy of prediction. Finally, the deep network yielded two output channels. The loss was calculated by cross-entropy cost function, and the final result was obtained by an 'argmax' function, in which each mesh face was classified as sulcus or gyrus (Equation(3)).

\begin{equation}\label{eq3}  
L=-[y \log \widehat{y}+(1-y) \log (1-\widehat{y})]
\end{equation}

The segmentation accuracy was defined in Equation(4), in which the ${N_f}$ denotes the number of mesh faces, $\mathbf{prd}_{i}$ and $\mathbf{lab}_{i}$ denotes the $i^{th}$ face's prediction and label, respectively. $sign$ function outputs 0 and 1, according to whether $\mathbf{prd}_{i}$ is equal to $\mathbf{lab}_{i}$ or not.

\begin{equation}\label{eq4}  
Acc =\frac{1}{N_f} \sum_{i=1}^{N_f}  sign(\mathbf{prd}_{i},\mathbf{lab}_{i})
\end{equation}
\section{Experiments}
\subsection{Dataset}

We used the 3-Tesla T1-weighted MRI and resting-state fMRI (rsfMRI) data from Human Connectome Project (HCP) dataset \cite{van2013wu}. These data have been minimally pre-processed upon release \cite{glasser2013minimal}. Important imaging parameters are: T1-weighted MRI: TR=2400 $ms$, TE=2.14 $ms$, flip angle=8 $deg$, image matrix=260×311×260 and resolution=0.7×0.7×0.7 $mm^3$. Diffusion MRI: TR=5520 $ms$, TE=89.5 $ms$, flip angle=78 $deg$, FOV=210×180 $mm^2$, matrix=168×144, resolution=1.25×1.25×1.25 $mm^3$, echo spacing=0.78 $ms$; rsfMRI: TR=0.72 $s$, TE=33.1 $ms$, flip angle=52 $deg$, in-plane FOV=208×180 $mm^2$, matrix=90×104, 220 $mm$ FOV, 72 slices, 1200 time points, 2.0 $mm$ isotropic voxels, BW = 2290 $Hz/Px$. 
T1-weighted MRI was used to reconstruct a cortical surface, since it provides more precise anatomical information due to its higher spatial resolution. Preprocessing of T1-weighted MRI includes skull stripping, tissue-segmentation and white matter surface reconstruction. The surfaces were transferred to the grayordinate system, a common surface template, \emph{via} cross-subject registration to standard volume and surface spaces. It is noted that the aligned surface of each subject was re-sampled such that vertices have cross-subject correspondence. Preprocessing of fMRI signals includes: ICA-denoise, spatial artifact/distortion removal. Within‐subject cross‐modal registration was applied such that each grayordinate vertex was associated with a preprocessed fMRI signal. More details of data acquisition and preprocessing are referred to \cite{glasser2013minimal}. Our experiments included 348 samples for training/testing (0.8/0.2).

\begin{figure}[!htp]
  \centering
  \includegraphics[width=10cm ]{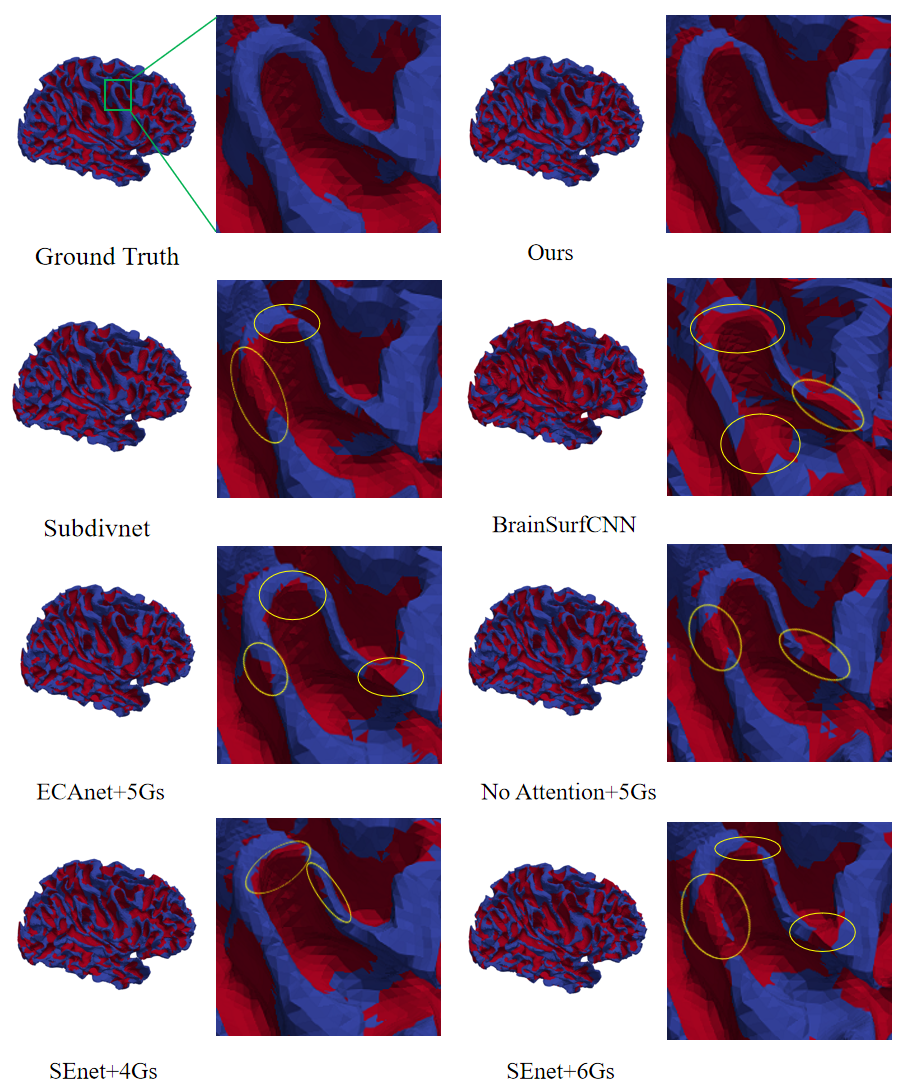}
  \caption{Comparison experiments between ground truth and predicated results \emph{via} different methods. A 'U' shaped gyral area is enlarged for inspection in detail, where yellow circles highlight the unsatisfied predictions. Abbreviations: G for gradients}
\end{figure}

\subsection{Implementation Details}

Firstly, to adapt to the requirement in Subdivnet, we re-meshed the original surface and transferred the gradient features as well as class labels (0 for gyri and 1 for sulci) from mesh vertices to the mesh faces. The batch size and learning rate was set to 6 and 0.02, respectively. 'Bilinear' upsample operation was used. The accuracy on testing samples converges (75.85\%) after about 20 epochs (from 15 to 21). Then, we chose the checkpoint of the best accuracy on test data set and continued fine-tuning for several epochs, with the learning rate set to 0.008.

\subsection{Comparison with Mesh Convolution Baselines and Other Models}

Since our model was modified from Subdivnet \cite{hu2022subdivision}, it was used as a mesh convolution baseline. The BrainSurfCNN \cite{jiang2019spherical} predicts the task fMRI contrasts from resting-state functional connectivity ones by using a regression scheme. For a fair comparison, we modified its' network architecture by replacing the 100 input channels with 5 channels, the 47 output channels with 2 channels in our model, as well as the loss function with cross-entropy loss, to implement the same segmentation task. The features on the mesh vertices were transferred to faces as well. A classic linear classification method, support-vector machine (SVM) \cite{hearst1998support} was also selected for comparison. Figure 2 shows the segmentation results from one random testing subject and Table 1 reports the comparison of results \emph{via} different methods by means of a variety of metrics. It is found that 
nonlinear models overwhelmingly outperform linear SVM in all metrics. Besides the intrinsic nonlinear relation between function and folding that could be significantly better encoded and decoded by nonlinear models, the definition of neighboring mask, a key factor for convolution operations on a mesh, could play an important role in the improvement. The importance of the convolution mask was also supported by a better performance of Subdivnet than BrainSurfCNN, since
BrainSurfCNN, unlike Subdivnet, only used a small neighborhood as convolution mask which limits the extraction of long-range features.
It is also found that sulci were often oversegmented by Subdivnet and BrainSurfCNN (yellow circles in Figure 2) in highly bent gyral regions, where the cortical morphology is far more complex than others.

\subsection{Ablation Study}
We considered the effects of different types of attention models and the number of functional gradient channels. It is found in Figure 2 and Table 2 that adding channel attention block or switching to another one does not significantly improve the performance in MIoU, Precision, F1 and Recall, whereas the accuracy could be affected by the selection of a channel attention block. The choice or absence of a channel attention could yield wrong segmentation in "straight" gyral/sulcal regions (the left yellow circles in third row of Figure 2). Neither fewer (4G) nor more (6G) functional gradients improve the segmentation performance. Mis-classification with 4G and 6G was usually found on the border between gyri and sulci (yellow circles in the fourth row of Figure 2).

\begin{table}[!h]
  \caption{Comparison with other methods. The MIoU, Precision, F1-score, Recall and Accuracy are reported.}
  \label{sample-table}
  \centering
  \begin{tabular}{llllll}
    \toprule
    
    Method     & MIoU     & Precision  & F1-score & Recall & Accuracy \\
    \midrule
    Our Method & 0.603  & 0.752  &0.751 &0.750 &75.85\%   \\
    Subdivnet     & 0.595 & 0.744   &0.745 &0.746	&75.11\%  \\
    BrainSurfCNN     & 0.464	&0.657	&0.622	&0.626	&65.03\%  \\
    Linear SVM     &0.295	&0.488	&0.371	&0.499	&59.09\% \\
    \bottomrule
  \end{tabular}
\end{table}

\begin{table}[!h]
  \caption{Ablation study on attention block and gradients. The MIoU, Precision, F1-score, Recall and Accuracy are reported. \textcolor{red}{Red} and \textcolor{blue}{blue} denote the best and the second-best results, respectively. Abbreviations: G for gradients.}
  \label{sample-table}
  \centering
  \begin{tabular}{lllllll}
    \toprule
    
    Type     & MIoU     & Precision  & F1-score & Recall & Accuracy \\
    \midrule
    SEnet+5Gs     &\textcolor{red}{0.603}	&\textcolor{red}{0.752}	&\textcolor{red}{0.751}	&\textcolor{red}{0.750}	&\textcolor{red}{75.85\%} 	 \\
    ECAnet+5Gs     &\textcolor{blue}{0.599} 	&\textcolor{blue}{0.750}	&\textcolor{blue}{0.747}	&0.745	&\textcolor{blue}{75.61\%}  \\
    No Attention+5Gs &0.598 &0.748 &0.747 &\textcolor{blue}{0.746} &75.45\% & \\
    SEnet+4Gs     &0.587  &0.744    &0.737 &0.734	&74.89\% \\
    SEnet+6Gs     &0.580 	&0.732	&0.732	&0.732	&73.86\%  \\
    \bottomrule
  \end{tabular}
\end{table}
\section{Discussion}
\begin{figure}[!htp]
  \centering
  \includegraphics[width=14cm]{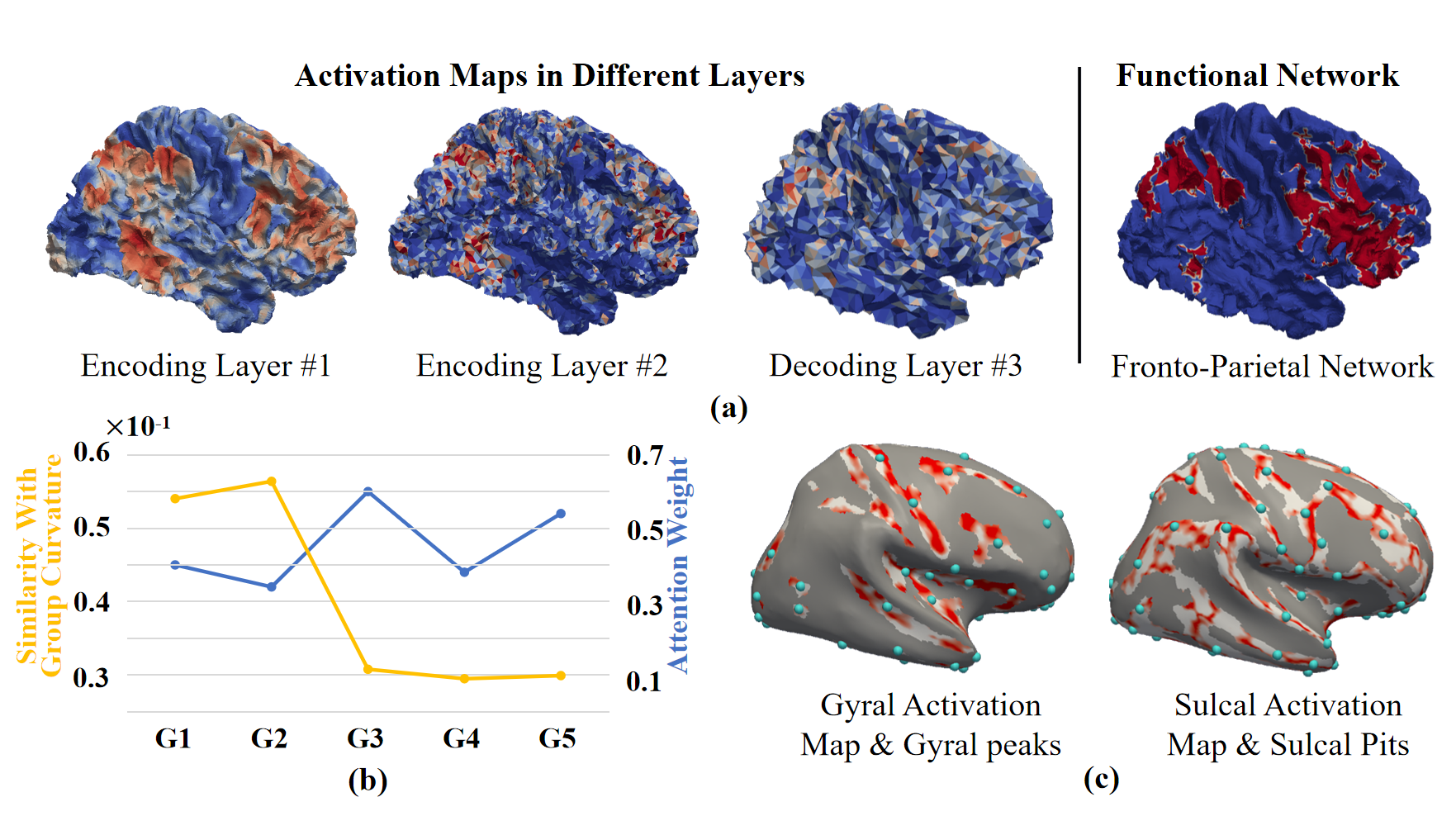}
  \caption{(a) Activation maps in different deep layers that resemble the fronto-parietal functional network, not detected by functional gradients. (b) Similarity between functional gradients (G for short) and curvature map, and the weights of Gs learnt from attention block. (c) The segmentation activation maps (red for more activated regions) in the last two-channel layer and cortical landmarks (cyan bubbles), including gyral peaks and sulcal pits, that are independent from curvature.}
\end{figure}

The attention block re-weights the five gradients. The latter gradients (3G and 5G) gain relatively higher weights (Figure 3(b)), and their similarity to curvature map (measured by Pearson correlation) is significantly lower than the dominant gradients (1G and 2G). These results suggest that the implicit relation between function and folding might not be encoded in dominant gradients. It is also interesting to find that some functional networks, which are not explicitly represented by input gradients, can be identified in activation maps in deep layers. As an example, the fronto-parietal network, one of the resting-state networks \cite{smith2013functional} not present in the 5 gradients, were found in both encoding and decoding layers, as shown in Figure 3(a). Finally, the activation maps in the last two-channel layer, were related to folding landmarks, gyral peaks (local convexity maxima) and sulcal pits (local concavity maxima) \cite{li2022cortical}, which were obtained independent of curvature used to yield the output gyro-sulcal maps. These landmarks (blue bubbles in Figure 3(c), group-wise mean locations across subjects) are not located on the highly activated regions (red regions). On average, activation values on these landmarks are 0.46±0.62 and 0.60±0.55 (\emph{p}<0.05, \emph{via} unpaired \emph{t}-test) on other regions (expect the gray background), suggesting that these landmarks might mark the borders of folding patterns that are highly correlated with brain functions.

\section{Conclusion}
In this work, we developed an attention mesh convolution model based on the U-net architecture to predict cortical gyro-sulcal segmentation maps from brain functional gradients derived from resting-state fMRI. The model architecture, such as input channel, output channel, feature types and attention block, can be easily extended or customized according to specific needs. Unlike the previous study, where functional signals from gyral/sulcal vertices were independently used as samples to a CNN model and their frequency band was found to be a discriminative feature \cite{liu2019cerebral}, the mesh convolution in our model considers the spatial organization of functional gradients and folding patterns on a cortical sheet. The prediction performance \emph{via} our model outperforms the state-of-the-art. In addition, we found that the dominant functional gradients contribute less to folding prediction. On the activation maps of the last layer, we found that some well-studied cortical landmarks are on the borders of, rather than within, the highly activated regions. These results and findings suggest that a specifically designed artificial neural network can improve the precision of the mapping between brain functions and cortical folding patterns, and can provide valuable clues of brain anatomy-function relation for neuroscience and artificial neural network design.

{
\small
\bibliographystyle{unsrt}
\bibliography{neurips_2022} 
}

\end{document}